\documentclass[11pt,aps,showpacs,showkeys,preprintnumbers,amsmath,amssymb,nofootinbib]{revtex4}

\usepackage{subfigure}
\usepackage{graphicx}
\usepackage{color}
\usepackage{amsmath}
\usepackage{amsfonts}
\usepackage{amssymb}
\usepackage{adjustbox}
\def \be  {\begin{equation}}
\def \ee  {\end{equation}}
\def \ee  {\end{equation}}
\def \bea {\begin{eqnarray}}
\def \eea {\end{eqnarray}}

\providecommand\textprime{}

\begin{document}

\vspace*{1mm}

\title{An approach of statistical corrections to interactions in hadron resonance gas}

\author{Mahmoud Hanafy}
\email{mahmoud.nasar@fsc.bu.edu.eg}
\affiliation{Physics Department, Faculty of Science, Benha University, 13518, Benha, Egypt}

\author{Muhammad Maher}
\email{m.maher@science.helwan.edu.eg} 
\affiliation{Helwan University, Faculty of Science, Physics Department, 11795 Ain Helwan, Egypt}

\date{\today}

\begin{abstract}

We propose a new model for hadrons with quantum mechanical attractive and repulsive interactions sensitive to some spatial correlation length parameter inspired by Beth-Uhlenbeck quantum mechanical non-ideal gas model \cite{uhlenbeck1937quantum}. We confront the thermodynamics calculated using our model with a corresponding recent lattice data at four different values of the baryon chemical potential, $\mu_{\mathtt{b}}= 0, 170, 340, 425~$MeV over temperatures ranging from $130$ MeV to $200~$MeV and for five values for the correlation length ranging from $0$ to $0.2~$fm. For equilibrium temperatures up to the vicinity of the chiral phase transition temperature $\simeq 160~$MeV, a decent fitting between the model and the lattice data is observed for different values of $r$, especially at 
$(\mu_{\mathtt{b}}, r) = (170,0.05), (340,0.1)$, and $(340,0.15)$, where $\mu_{\mathtt{b}}$ is in MeV and $r$ is in fm. For vanishing chemical potential, the uncorrelated model ($r=0$), which corresponds to ideal hadron resonance gas model seems to offer the best fit. The quantum hadron correlations seem to be more probable at non-vanishing chemical potentials, especially within the range $\mu_{\mathtt{b}}\in [170, 340~$MeV$]$.

\end{abstract}


\maketitle

\section{Introduction}
\label{sec:Intr}
  
One of the key goals of the ultrarelativistic nuclear collisions is to gain information on the hadron-parton phase diagram, which is characterized by different phases and different types of the phase transitions \cite{banks1983deconfining}. Quantum Chromodynamics (QCD), the gauge field theory that describes the strong interactions of colored quarks and gluons and their colorless bound states, has two important intensive state parameters at equilibrium, namely temperature $T$ and baryon chemical potential $\mu_{\mathtt{b}}$. A remarkable world-wide theoretical and experimental effort has been dedicated to the study of strongly interacting matter under extreme condition of temperature and baryon chemical potential. The lattice QCD simulations provide an \textit{a priori} non-perturbative regularization of QCD that makes it compliant with analytic and computational methods with no model assumptions other than QCD itself being needed to formulate the theory. The temperature and density (chemical potential) dependence of the bulk thermodynamic quantities, commonly summarized as the equation of state (EoS), provides the most basic characterization of equilibrium properties of the strongly interacting matter. Its analysis within the framework of lattice QCD has been refined ever since the early calculations performed in pure $SU(N)$ gauge theories \cite{engels1981high}. The EoS at vanishing chemical potentials does already provide important input into the modeling of the hydrodynamic evolution of hot and dense matter created in heavy-ion collisions \cite{Karsch:2003zq,Tawfik:2004sw}. While this is appropriate for the thermal conditions met in these collisions at the LHC and the highest RHIC beam energies, knowledge of the EoS at non-vanishing baryon, strangeness and electric charge chemical potentials is indispensable for the hydrodynamic models of the conditions met in the beam energy scan (BES) at RHIC \cite{Gyulassy:2004zy} and in future experiments at facilities like FAIR at GSI and NICA at JINR \cite{Tawfik:2016cot,Tawfik:2016wfm}. 

Bulk thermodynamic observables such as pressure, energy density, and entropy density as well as the second order quantities such as the specific heat and velocity of sound have now been obtained at vanishing chemical potentials for the three lightest quark flavors \cite{Tawfik:2004vv}. By the analysis of the chiral transition temperature, $T_{\mathtt{c}}\simeq 154 \pm 9$ MeV \cite{bazavov2012chiral} it has also been shown that the bulk thermodynamic observables change smoothly in the transition region probed. Due to the well-known sign problem encountered in the lattice QCD formulations at finite chemical potential, a direct calculation of the EoS at non-zero chemical potential is unfortunately not fully reliable. A lot of efforts has been made to circumvent the divergences at non-zero chemical potential, such as Taylor expansion of the thermodynamic potential on course lattices besides other sophisticated computational techniques \cite{gavai2001quark,allton2002qcd,ejiri2006isentropic,borsanyi2012qcd,gunther2017qcd,d2017higher} which made it possible to conduct calculations covering the range $0 \leq \mu_{\mathtt{b}}/T \lesssim 3$ that is expected to be explored with the BES at RHIC by varying the beam energies in the range $7.7\leq \sqrt{s_{\mathtt{NN}}} \leq 200~$GeV \cite{Bzdak:2019pkr}. A promising approach in this quest is the investigation of hadron production. The hadron resonance gas (HRG) is customarily used in the lattice QCD calculations as a reference for  the hadronic sector \cite{Huovinen:2009yb,Borsanyi:2010bp}. At low temperatures, they are found to be in quite good agreement with the HRG model calculations \cite{Tawfik:2014eba}, although some systematic deviations have been observed, which may be attributed to the existence of additional resonances which are not taken into account in HRG model calculations based on well established resonances listed by the particle data group \cite{Tanabashi:2018oca} and perhaps to need to extend the model to incorporate interactions. 

In the HRG model, the thermodynamics of a strongly interacting system is conjectured to be approximated as an ideal gas composed of hadron resonances with masses $\lesssim2~$GeV \cite{Tawfik:2014eba,Karsch:2003zq} that are treated as a free gas, exclusively in the hadronic phase, i.e. below $T_c$. Therefore, the hadronic phase in the confined phase of QCD could be modeled as a non-interacting gas of the hadron resonances. It is reported in recent literature that the standard performance of the HRG model seems to be unable to describe all the available data that is predicted by recent lattice QCD simulations \cite{samanta2019exploring,vovchenko2017multicomponent}. The conjecture to incorporate various types of interactions has been worked out in various studies \cite{Tawfik:2004sw,Huovinen:2017ogf,Vovchenko:2017xad,Vovchenko:2017drx}. When comparing the thermodynamics calculated within the HRG framework with the corresponding data obtained using lattice QCD methods, one has to decide how to incorporate interactions among the hadrons.

Arguments based on the S-matrix approach \cite{dashen1969s,venugopalan1992thermal,lo2017s} suggest that the HRG model includes attractive interactions between hadrons which lead to the formation of resonances. More realistic hadronic models take into account the contribution of both attractive and repulsive interactions between the component hadrons. Repulsive interactions in the HRG model had previously been considered in the framework of the relativistic cluster and virial expansions \cite{venugopalan1992thermal}, via repulsive mean fields \cite{olive1981thermodynamics,olive1982quark}, and via excluded volume (EV) corrections \cite{hagedorn1980hot,gorenstein1981phase,hagedorn1983pressure,kapusta1982thermodynamics,
rischke1991excluded,anchishkin1995generalization}. In particular, the effects of EV interactions between hadrons on HRG thermodynamics \cite{satarov2009equation,andronic2012interacting,bhattacharyya2014fluctuations,
albright2014matching,vovchenko2015hadron,albright2015baryon,redlich2016thermodynamics,
alba2017excluded} and on observables in heavy-ion collisions \cite{braun1999chemical,cleymans2006comparison,begun2013hadron,fu2013higher,
vovchenko2017examination,alba2018flavor,satarov2017new,vovchenko2017van} has extensively been studied in the literature. Recently, repulsive interactions have received renewed interest in the
context of lattice QCD data on fluctuations of conserved charges. It was shown that large deviations of several fluctuation observables from the ideal HRG baseline could well be interpreted
in terms of repulsive baryon-baryon interactions \cite{huovinen2018hadron,vovchenko2017van,vovchenko2017equations}.

The present script is organized as follows: In Section \label{sec:Mod}, we review the detailed formalism of the conventional ideal (uncorrelated) HRG model, then we develop a non-ideal (correlated) statistical correction to the ideal HRG model inspired by Beth-Uhlenbeck (BU) quantum theory of non-ideal gases. The calculations of the HRG thermodynamics based on the proposed correction are discussed in Section \label{sec:Res}. Section \label{sec:Cncls} is devoted to the conclusions and outlook.

\section{Model Description}
\label{sec:Mod}

In this study, we use the particle interaction probability term originally implemented in the expression for the second virial coefficient worked out in ref. \cite{uhlenbeck1937quantum} in order to suggest a statistical correction to the uncorrelated HRG model. Beth and Uhlenbeck suggested a connection between the virial coefficients and the probabilities of finding pairs, triples and so on, of particles near each other \cite{uhlenbeck1936quantum}. In the classical limit, which is usually designated by sufficiently high temperatures and/or low particle densities, it was shown that these probabilities (explicit expressions are to follow next section) can be expressed by Boltzmann factors so long as the de Broglie wavelength, which is a common measure of the significance of the quantum non-localiy, is small enough compared with the particle spacial extent measured by the particle {''diameter''} \cite{uhlenbeck1936quantum}. Such a particle diameter can be considered as a measure of the spatial extent within which a particle can undergo hardcore (classical) interactions. Based on a comparison of model with experimental results, it was concluded  that at sufficiently low temperatures for which the thermal de Broglie wave length is comparable with the particle diameter, deviations from the classical excluded volume model due quantum effects will be significant \cite{uhlenbeck1936quantum}. 
 
An extension has been made to the quantum mechanical model of the particle interactions as proposed in ref. \cite{uhlenbeck1936quantum} by considering the influence of Bose or Fermi statistics in addition to the effect of the inclusion of discrete quantum states for a general interaction potential that is not necessarily central \cite{uhlenbeck1937quantum}. The expression for the second virial expansion developed in ref. \cite{uhlenbeck1936quantum} and then extended in ref. \cite{uhlenbeck1937quantum} was later generalized using the cluster integral to describe particle interactions provided that those particles don't form bound states \cite{dashen1969s,venugopalan1992thermal,kostyuk2001second}.
 
The quantum mechanical Beth-Uhlenbeck (BU) approach were quite recently used to model the repulsive interactions between baryons in a hadron gas \cite{vovchenko2018beth}. The second virial coefficient or the “excluded volume” parameter was calculated within the BU approach and found to be temperature dependent, and found also to differ dramatically from the classical excluded volume (EV) model result. At temperatures $T=100-200$ MeV, the widely used classical EV model \cite{huang1963statistical,landau1980lifshitz,greiner2012thermodynamics} underestimates the EV parameter for nucleons at a given value of the nucleon hard-core radius (assumed $\simeq 0.3$ fm) by large factors of 3-4. It was thus concluded in \cite{vovchenko2018beth} that previous studies, which employed the hard-core radii of hadrons as an input into the classical EV model, have to be re-evaluated using the appropriately rescaled quantum mechanical EV parameters.

In this section, we first introduce the basic formulation of the ideal HRG model. Then we develop a statistical model inspired by Beth-Uhlenbeck (BU) quantum theory of non-ideal (correlated) gases \cite{uhlenbeck1937quantum} as a correction to the ideal (uncorrelated) HRG model. We thereby implement in our calculations a modified version of the partition function of a typical ideal gas. In the framework of bootstrap picture \cite{fast1963statistical,fast1963nuovo,eden1966polkinghorne}, an equilibrium thermal model for an interaction free gas has a partition function $Z(T, \mu, V)$ from which the thermodynamics of such a system can be deduced by taking the proper derivatives. 

\subsection{Non-correlated ideal HRG}

In a grand canonical ensemble, the partition function reads \cite{Tawfik:2014eba,Karsch:2003vd,Karsch:2003zq,Redlich:2004gp,Tawfik:2004sw,Tawfik:2005qh}
\bea 
Z(T,V,\mu)=\mbox{Tr}\left[\exp\left(\frac{{\mu}N-H}{T}\right)\right], \label{eq:Z} 
\eea
where $H$ is Hamiltonian combining all relevant degrees of freedom and $N$ is the number of constituents of the statistical ensemble. Eq. (\ref{eq:Z}) can be expressed as a sum over all hadron resonances taken from recent particle data group (PDG) \cite{Tanabashi:2018oca} with masses up to $2.5~$ GeV,
\bea 
\ln  Z(T,V,\mu)=\sum_i{{\ln Z}_i(T,V,\mu)} = V \sum_i \frac{g_i}{2{\pi}^2}\int^{\infty}_0{\pm p^2 dp {\ln} {\left[1\pm {\lambda}_i \exp\left(\frac{-{\varepsilon}_i(p)}{T} \right) \right]}}, \label{eq:lnZ}
\eea
where the pressure can be derived as $T\partial \ln  Z(T,V,\mu)/\partial V$, $\pm$ stands for fermions and bosons, respectively. $\varepsilon_{i}=\left(p^{2}+m_{i}^{2}\right)^{1/2}$ is the dispersion relation and $\lambda_i$ is the fugacity factor of the $i$-th particle \cite{Tawfik:2014eba},
\bea
\lambda_{i} (T,\mu)=\exp\left(\frac{B_{i} \mu_{\mathtt{b}}+S_{i} \mu_{\mathtt{S}}+Q_{i} \mu_{\mathtt{Q}}}{T} \right), \label{eq:lmbd}
\eea
where $B_{i} (\mu_{\mathtt{b}})$, $S_{i} (\mu_{\mathtt{S}})$, and $Q_{i} (\mu_{\mathtt{Q}})$ are baryon, strangeness, and electric charge quantum numbers (their corresponding chemical potentials) of the $i$-th hadron, respectively. From phenomenological point of view, the baryon chemical potential $\mu_{\mathtt{b}}$ - along the chemical freezeout boundary, where the production of particles is conjectured to cease - can be related to the nucleon-nucleon center-of-mass energy $\sqrt{s_{\mathtt{NN}}}$ \cite{tawfik2015thermal}
\bea
\mu_{\mathtt{b}} &=& \frac{a}{1+b \sqrt{s_{\mathtt{NN}}}}, \label{eq:mue}
\eea
where
$a=1.245\pm0.049~$GeV and $b=0.244\pm0.028~$GeV$^{-1}$. In addition to pressure, the number and energy density, respectively, and likewise the entropy density and other thermodynamics can straightforwardly be derived from the partition function by taking the proper derivatives
\begin{eqnarray}
n_i(T,\mu) &=& 
\sum_{i}\frac{g_{i}}{2\pi^{2}}\int_{0}^{\infty}{p^{2} dp \frac{1}{\exp\left[\frac{\mu_{i} - \varepsilon_{i}(p)}{T}\right] \pm 1}}, \\ 
\rho_i(T,\mu) &=& 
\sum_{i}\frac{g_{i}}{2\pi^{2}}\int_{0}^{\infty}{p^{2} dp \frac{-\varepsilon_{i(p)}\pm \mu_i}{\exp\left[\frac{\mu_{i}-\varepsilon_i(p)}{T}\right] \pm 1}}. \label{eq:e}
\end{eqnarray}

It should be noticed that both  $T$ and $\mu=B_{i} \mu_{\mathtt{b}}+S_{i} \mu_{\mathtt{S}}+\cdots$ are related to each other and to $\sqrt{s_{\mathtt{NN}}}$ \cite{Tawfik:2014eba}. As an overall thermal equilibrium is assumed, $\mu_{\mathtt{S}}$ is taken as a dependent variable to be estimated due to the strangeness conservation, i.e. at given $T$ and $\mu_{\mathtt{b}}$, the value assigned to $\mu_{S}$ is the one assuring $\langle n_{\mathtt{S}}\rangle-\langle n_{\bar{\mathtt{S}}}\rangle=0$. Only then, $\mu_{\mathtt{S}}$ is combined with $T$ and $\mu_{\mathtt{b}}$ in determining the thermodynamics, such as the particle number, energy, entropy, etc. The chemical potentials related to other quantum charges, such as the electric charge and the third-component isospin, etc. can also be determined as functions of $T$, $\mu_{\mathtt{b}}$, and $\mu_{\mathtt{S}}$ and each of them must fulfill the corresponding conservation laws.  

\subsection{Quantum-statistically correlated HRG}

As introduced in the previous section, for a quantum gas of fermions and bosons with mass $m_{i}$ and correlation (interaction) distance $r$, at temperature $T$ and vanishing $\mu_{\mathtt{b}}$, a two-particle interaction probability of the form 
\bea
1\pm \exp\left(-4\pi^{2} m_{i} T r^{2}\right)
\eea 
was first introduced by Beth and Uhlenbeck \cite{uhlenbeck1937quantum} in an attempt to model the interactions of a quantum gas of particles assuming a general potential and neglecting the possibility for bound states formation. The Boltzmann-like term $\exp\left(-4\pi^{2} {m}_{i} T r^{2}\right)$ remains in effect even for an ideal gas, which is a typical approximation at sufficiently high temperatures. The $\pm$ sign expresses the apparent attraction (repulsion) between bosons (fermions) due to change of statistics \cite{uhlenbeck1937quantum}. Inspired by such a correction, we introduce a correction for the probability term in the expression for the the ideal hadron gas partition function given in Eq. (\ref{eq:Z}). We propose a new probability term of the form
\bea
1\pm \lambda_i \exp\left(\frac{-{\varepsilon}_i(p)}{T}\right) \left[1\pm \exp\left(-4\pi^{2} {m}_{i} T r^{2}\right)\right].
\eea 


This corrected probability function obviously incorporates interactions in the the hadron resonance gas in the sense of Beth and Uhlenbeck quantum correlations \cite{uhlenbeck1937quantum} with $r$ being the  correlation (interaction) length between any two hadrons at equilibrium temperature $T$. Based on our proposed corrected probability function, we modify the non-correlated HRG partition function $Z(T, \mu, V)$ to have the following form   
\bea
\ln Z ^{\prime}  \textprime(T,V,\mu) = 
\sum_i V \frac{g_i}{2 {\pi}^2}\int^{\infty}_0{\pm p^2 dp {\ln} \left[1 \pm \lambda_{i} \exp\left(\frac{-{\varepsilon}_{i(p)}}{T} \right)  \left[1\pm \exp\left(4 \pi^{2} m_{i} T r^{2}\right)\right]\right]},  \label{eq:lnZ/} 
\eea
which apparently sums over all hadron resonances following the same recipe described in motivating Eq. (\ref{eq:lnZ}) for the case of non-correlated HRG. The thermodynamics of the correlated HRG can thus be calculated by taking the proper derivatives of $\ln Z \textprime$ as explicitly stated in the corresponding  non-correlated HRG case discussed above.

\section{Calculation Results}
\label{sec:Res}
  
We confront the data of the thermodynamics calculated using our statistically corrected HRG model based on Eq. (\ref{eq:lnZ/}) with the corresponding lattice thermodynamics data from \cite{bazavov2017qcd,bazavov2014equation} in the temperature range $T\in [130, 200~$MeV$]$. These temperatures are rather typical for the phenomenological applications in the context of heavy-ion collisions and lattice QCD equation-of-state. In refs. \cite{bazavov2017qcd,bazavov2014equation}, the authors calculated the QCD equation of state using Taylor expansions that include contributions from up to sixth order in the baryon, strangeness and electric charge chemical potentials. Calculations have been performed with a highly improved staggered quark action in the temperature range $T\in [130, 330~$MeV$]$ using up to four different sets of lattice cut-offs. The lattice data we are confronting to our model are shown on Figure 7. of ref. \cite{bazavov2017qcd}, (Left) the total pressure in (2+1)-flavor QCD and (Right) the total energy density in (2+1)-flavor QCD for several values of $\mu_{b}/T$.
  
\begin{figure}[!htb]
\includegraphics[width=8.25cm]{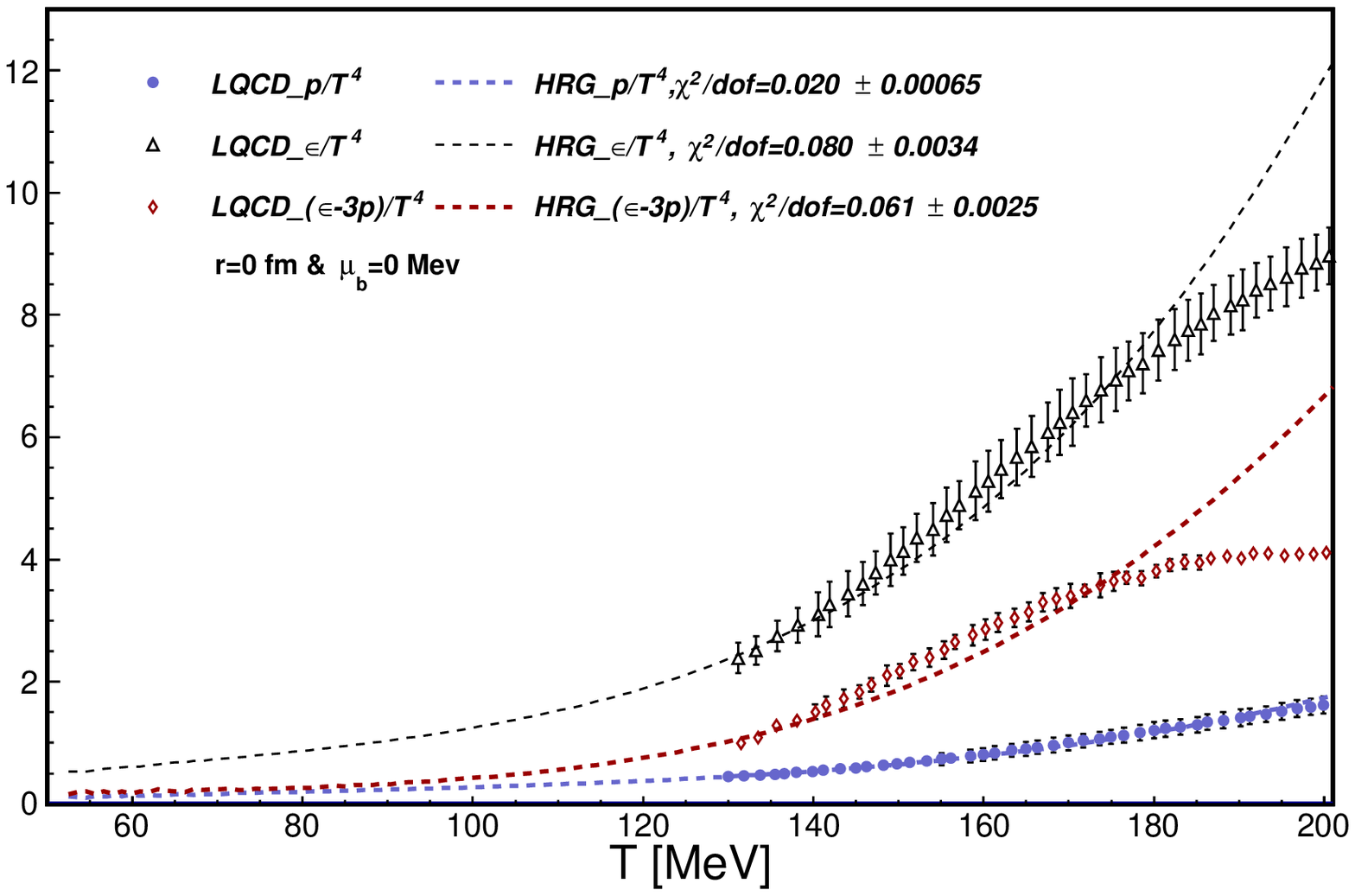}
\includegraphics[width=8.25cm]{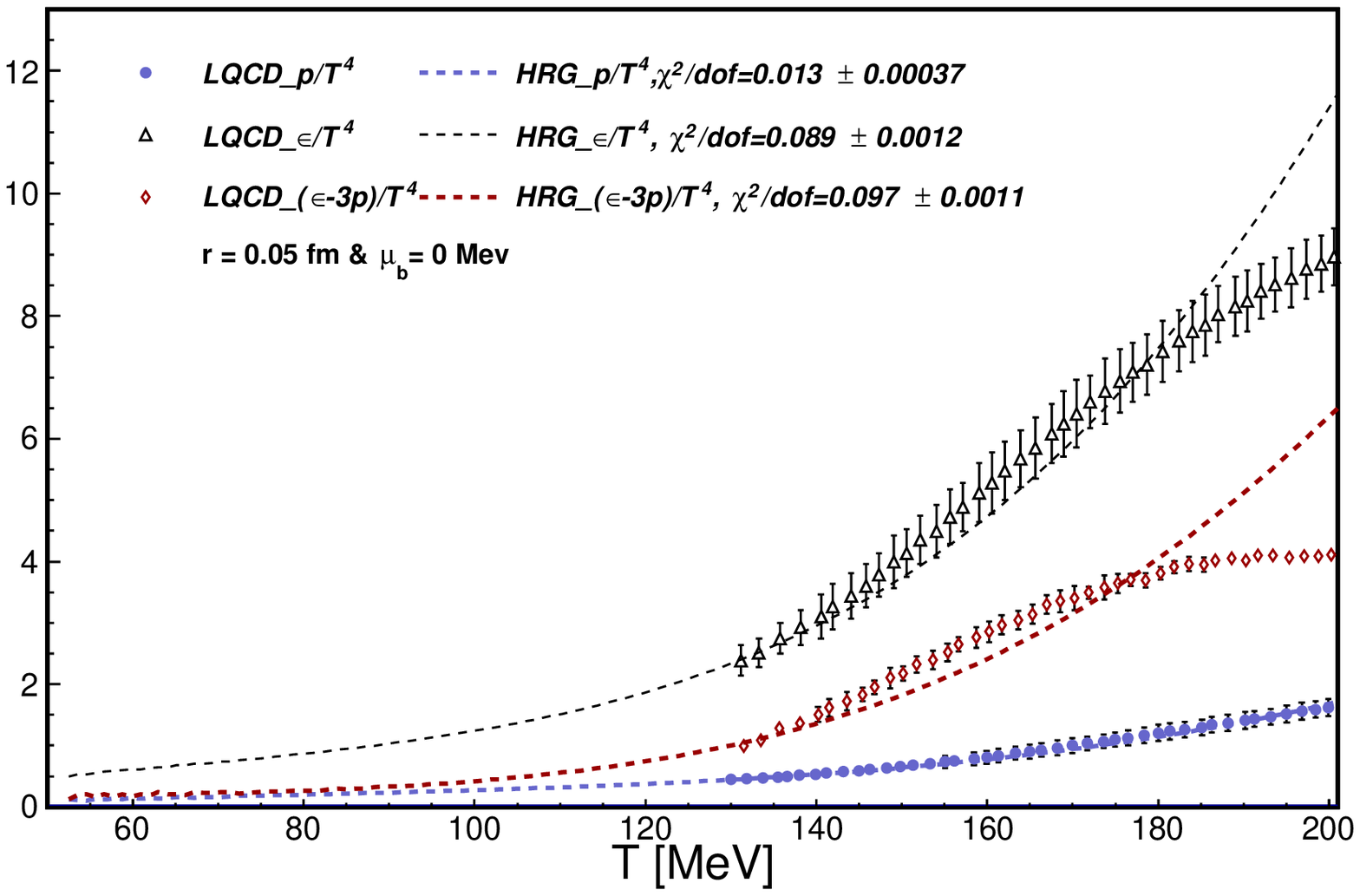} 
\includegraphics[width=8.25cm]{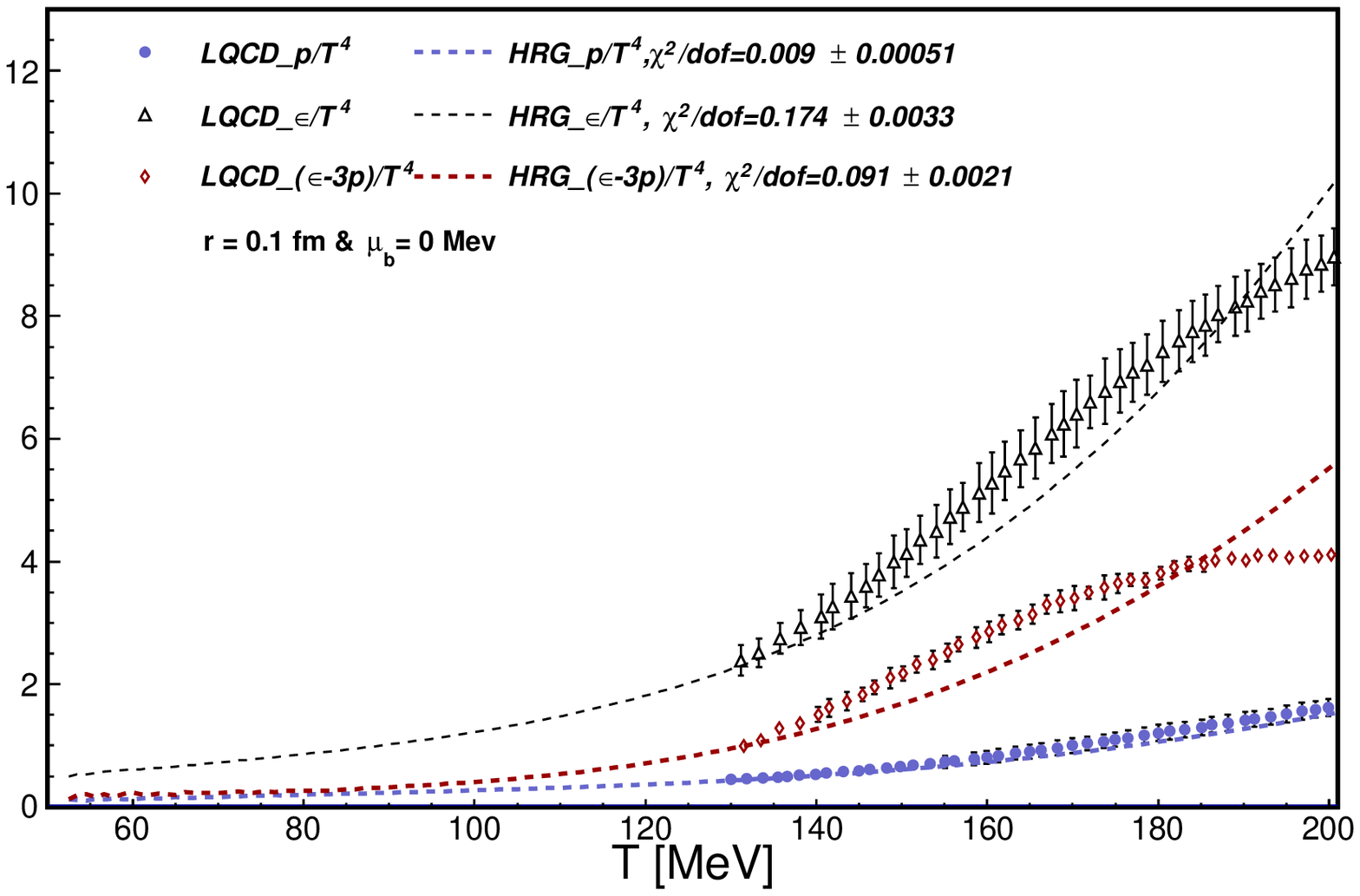} 
\includegraphics[width=8.25cm]{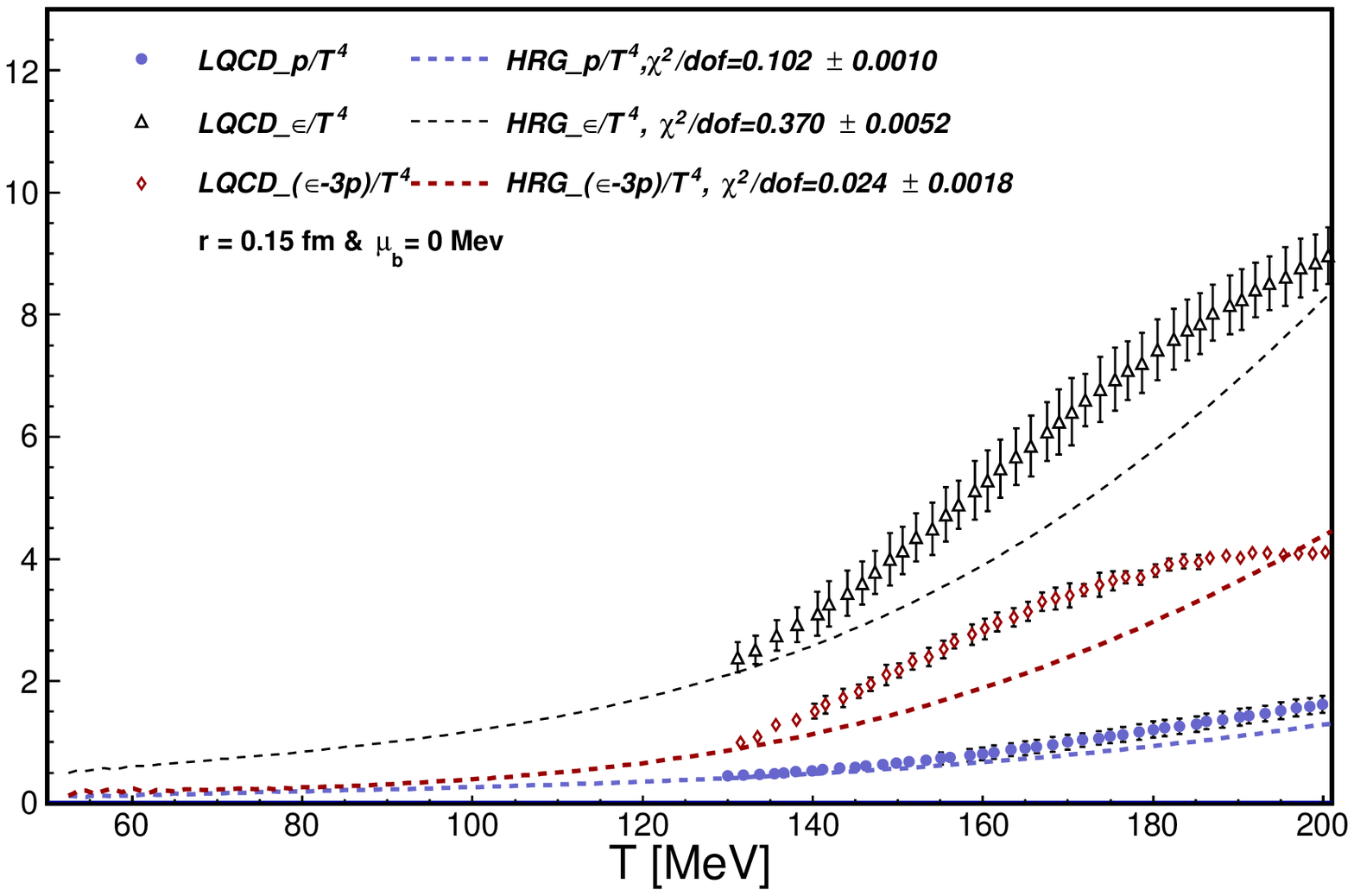}
\caption{Normalized pressure $P/T^{4}$, normalized energy density $\rho/T^{4}$ and trace anomaly $\left(\rho-3P\right)/T^{4}$ (dashed curves) calculated using our statistically corrected HRG model and confronted to the corresponding lattice data taken from ref. \cite{bazavov2017qcd} (symbols with error bars), at $\mu_{b}=0~$MeV. Comparison is made for four different values of the correlation length $r$.}
\label{fig:one}
\end{figure}

  Figure \ref{fig:one} of this letter depicts the temperature dependence of the normalized pressure $P/T^{4}$, normalized energy density $\rho/T^{4}$, and trace anomaly $\left(\rho-3P\right)/T^{4}$ (dashed curves) calculated using our statistically corrected HRG model based on Eq. (\ref{eq:lnZ/}). Moreover, in Figure \ref{fig:one}, our model data are confronted to the corresponding lattice data taken from ref. \cite{bazavov2017qcd} (symbols with error bars) at vanishing baryon chemical potential $\mu_{b}=0 MeV$. Comparison is made for four different values of the correlation length $r$. Table \ref{tab1} lists $\chi^{2}$/dof statistic for the normalized pressure $P/T^{4}$, trace anomaly $\left(\rho-3P\right)/T^{4}$ and normalized energy density $\rho/T^{4}$ calculated in our statistically corrected hadron resonance gas (HRG) model and confronted to the corresponding lattice data from \cite{bazavov2017qcd} for four values of baryon chemical potential $\mu_{b}=0,170, 340$, and $425~$MeV.   
  
At vanishing chemical potential, the best fit to the lattice data occurs for the case of zero correlation length which, in this case, corresponds to ideal HRG model. However, a slight exaggeration ($\chi^{2}/dof$ $\simeq$ 0.06 in the trace anomaly data) of our model's thermodynamics is observed in the vicinity of the critical phase transition temperature ($ T_{c}$ $\backsimeq$ 160 MeV). In the range $T\in [130, 200~$MeV$]$, the discrepancy between our model thermodynamics and the corresponding  lattice data is amplified. Generally, it is obvious that increasing the correlation length emphasizes the mismatch between our model and lattice data.

\begin{figure}[!htb]
\includegraphics[width=8.25cm]{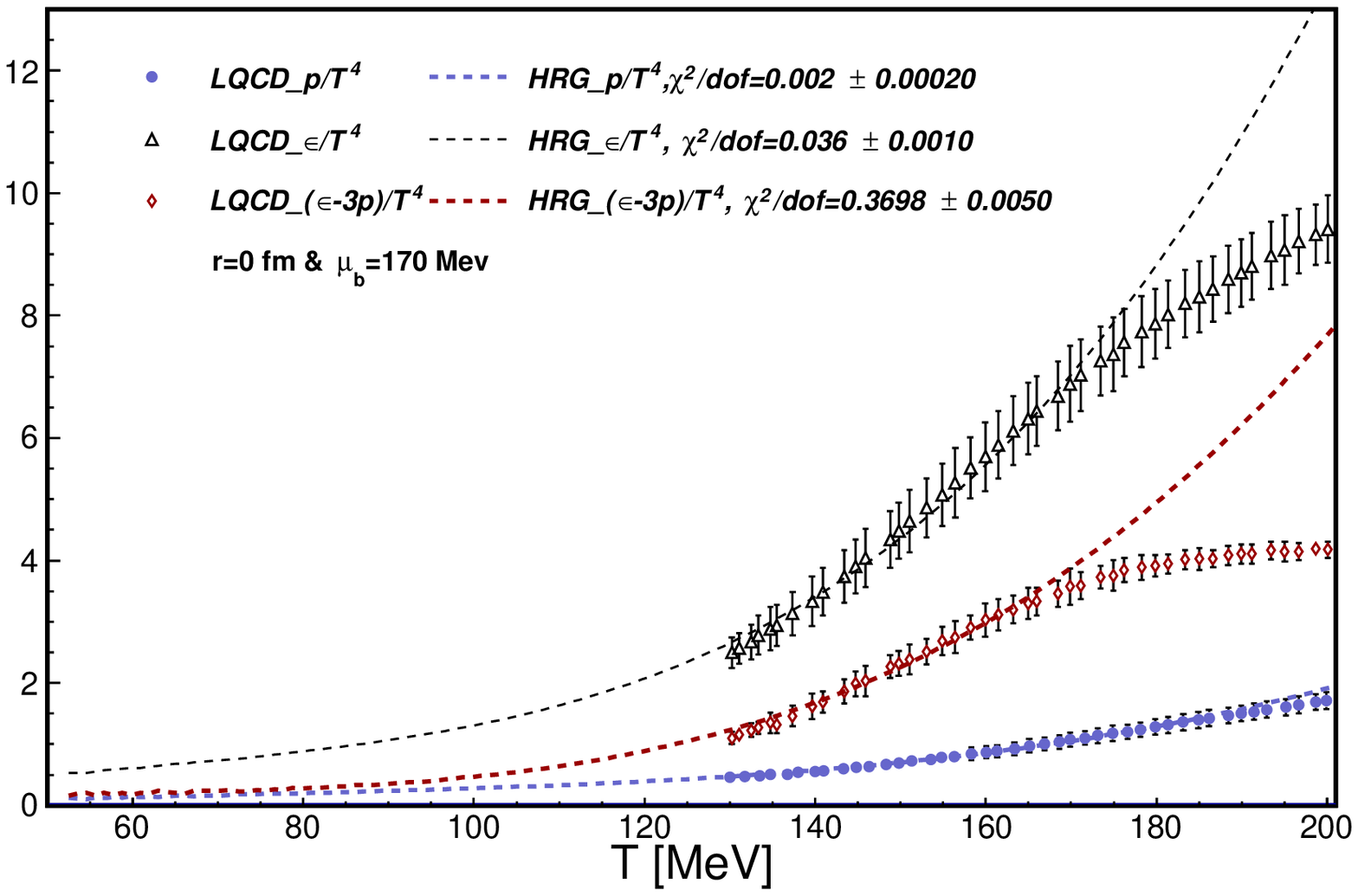}
\includegraphics[width=8.25cm]{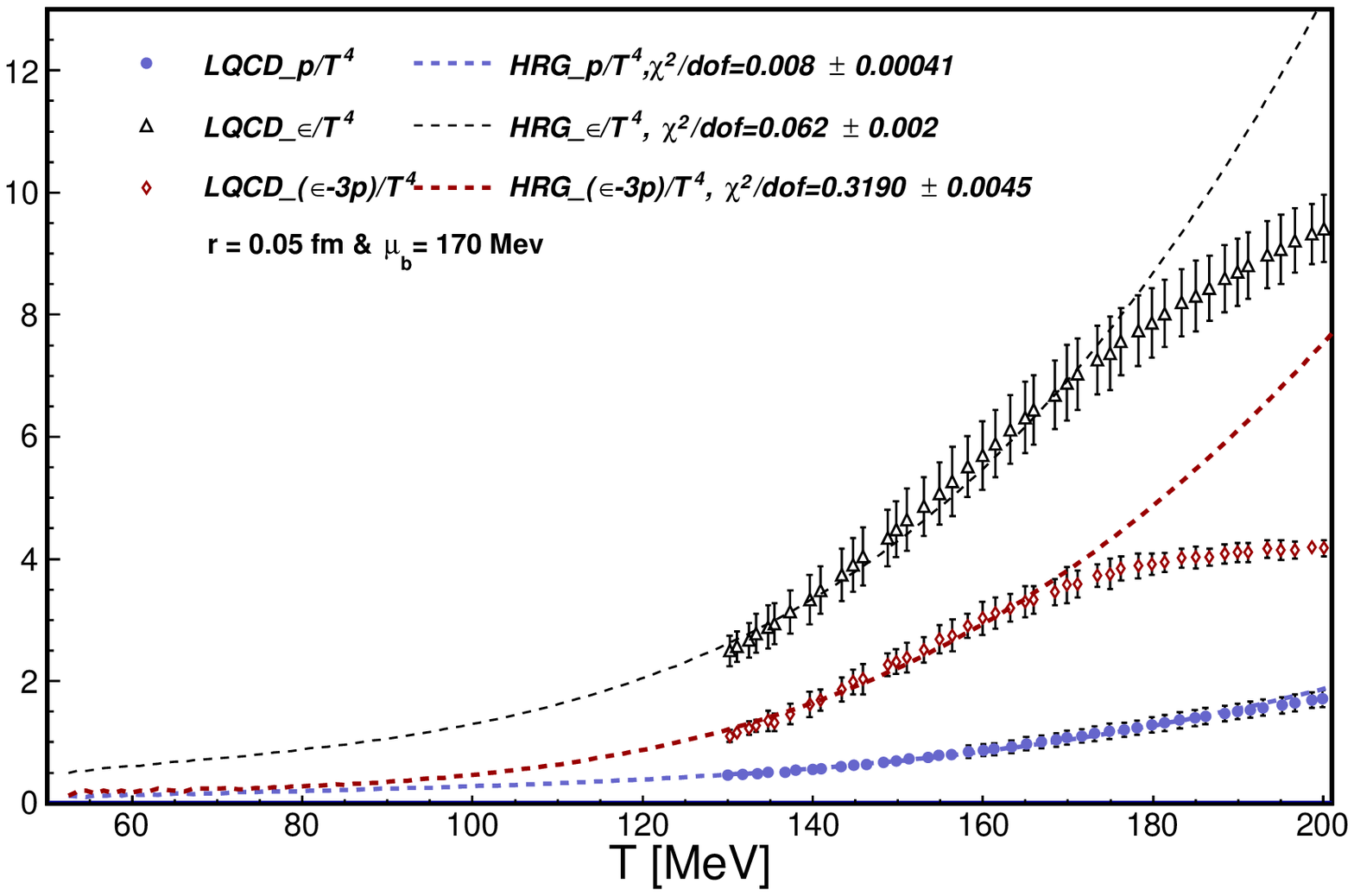} 
\includegraphics[width=8.25cm]{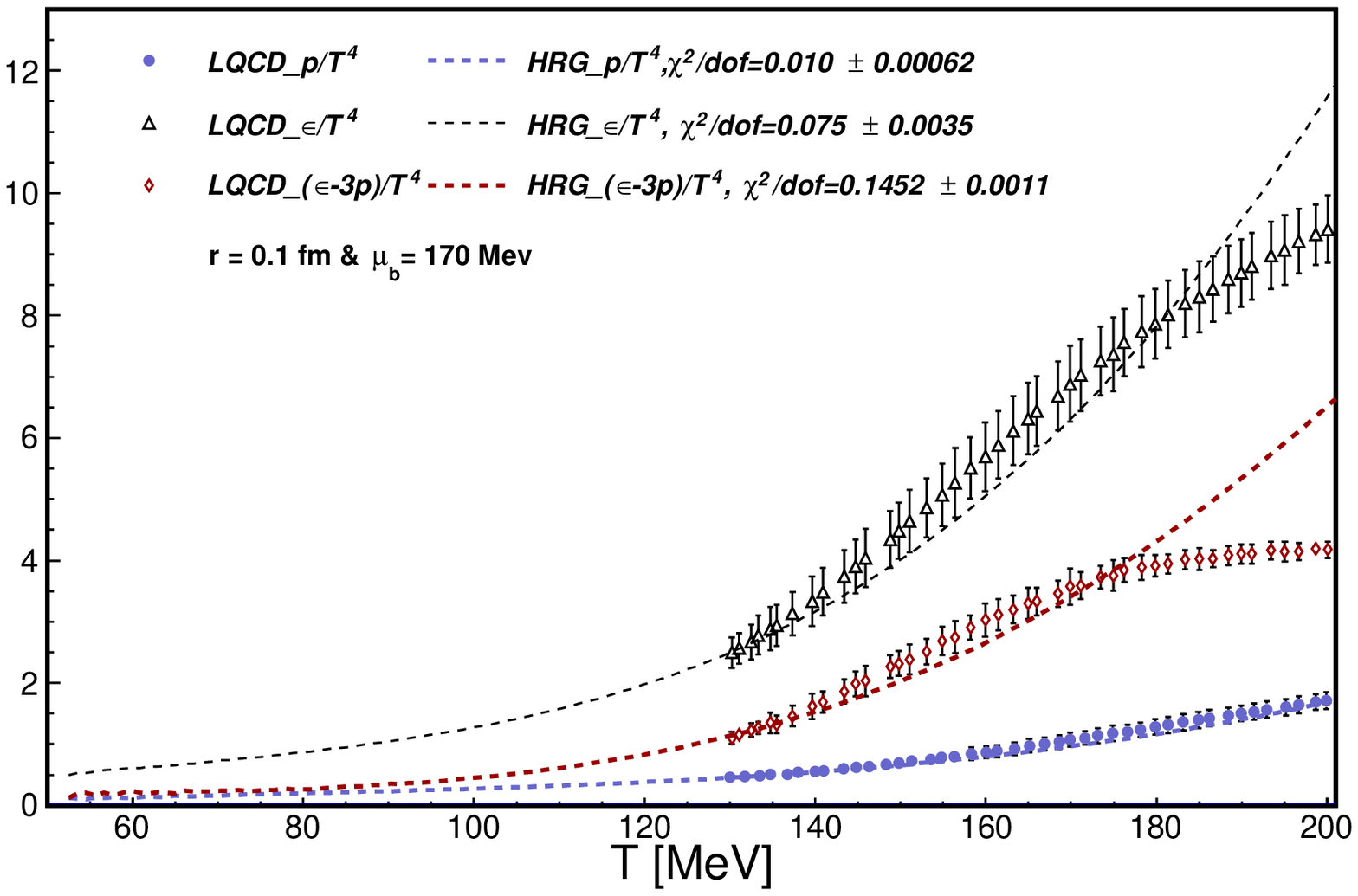} 
\includegraphics[width=8.25cm]{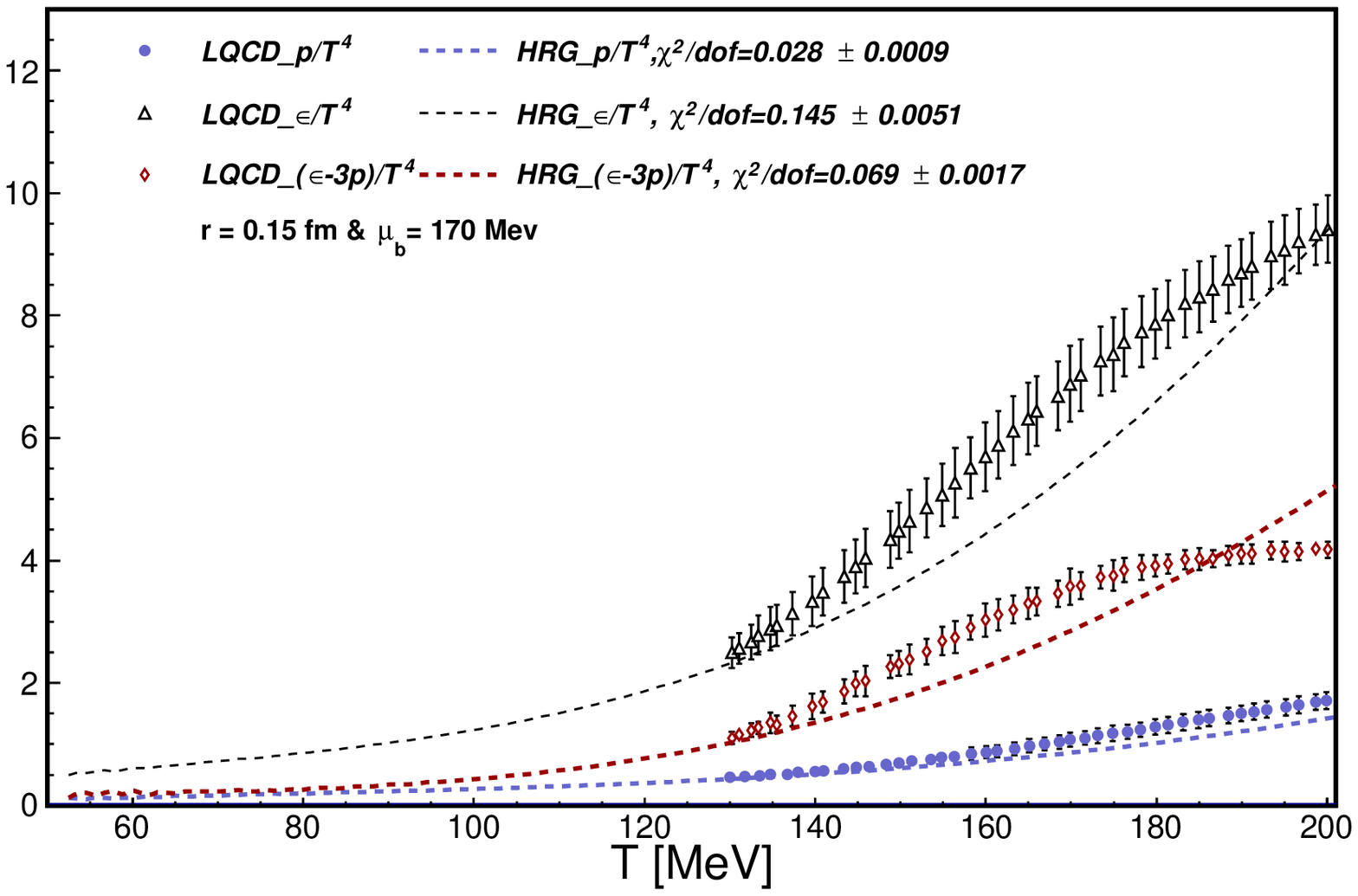}
\caption{The same as in Fig. \ref{fig:one} but at $\mu_{b}=170~$MeV.}
\label{fig:two}
\end{figure}

For the case $\mu_{b}=170~$MeV and as it appears in Fig. \ref{fig:two}, the best fit generally occurs for $r=0$ and for $r=0.05~$fm. This good-fit temperature range extends from temperatures well below $T_{c}$ till $T \gtrsim T_{c} \backsimeq 160~$MeV. It is quite obvious here that the model fits the lattice data better compared to the corresponding vanishing chemical potential case(s). However, in the temperature range $T\in [170, 200~$MeV$]$, the mismatch of our model with the lattice data becomes more pronounced compared to the corresponding range of the vanishing chemical potential case(s).

\begin{figure}[!htb]
\includegraphics[width=8.25cm]{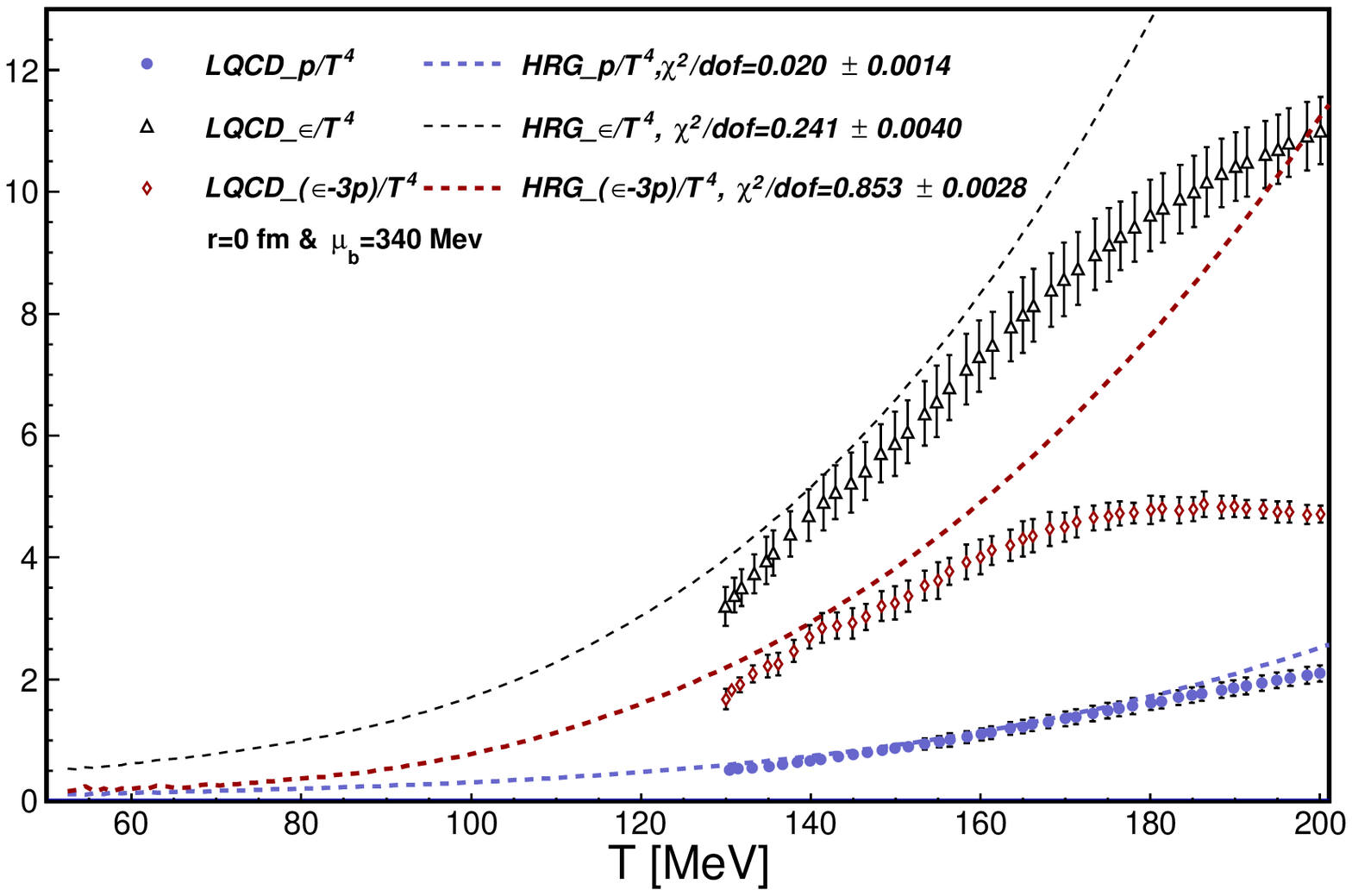}
\includegraphics[width=8.25cm]{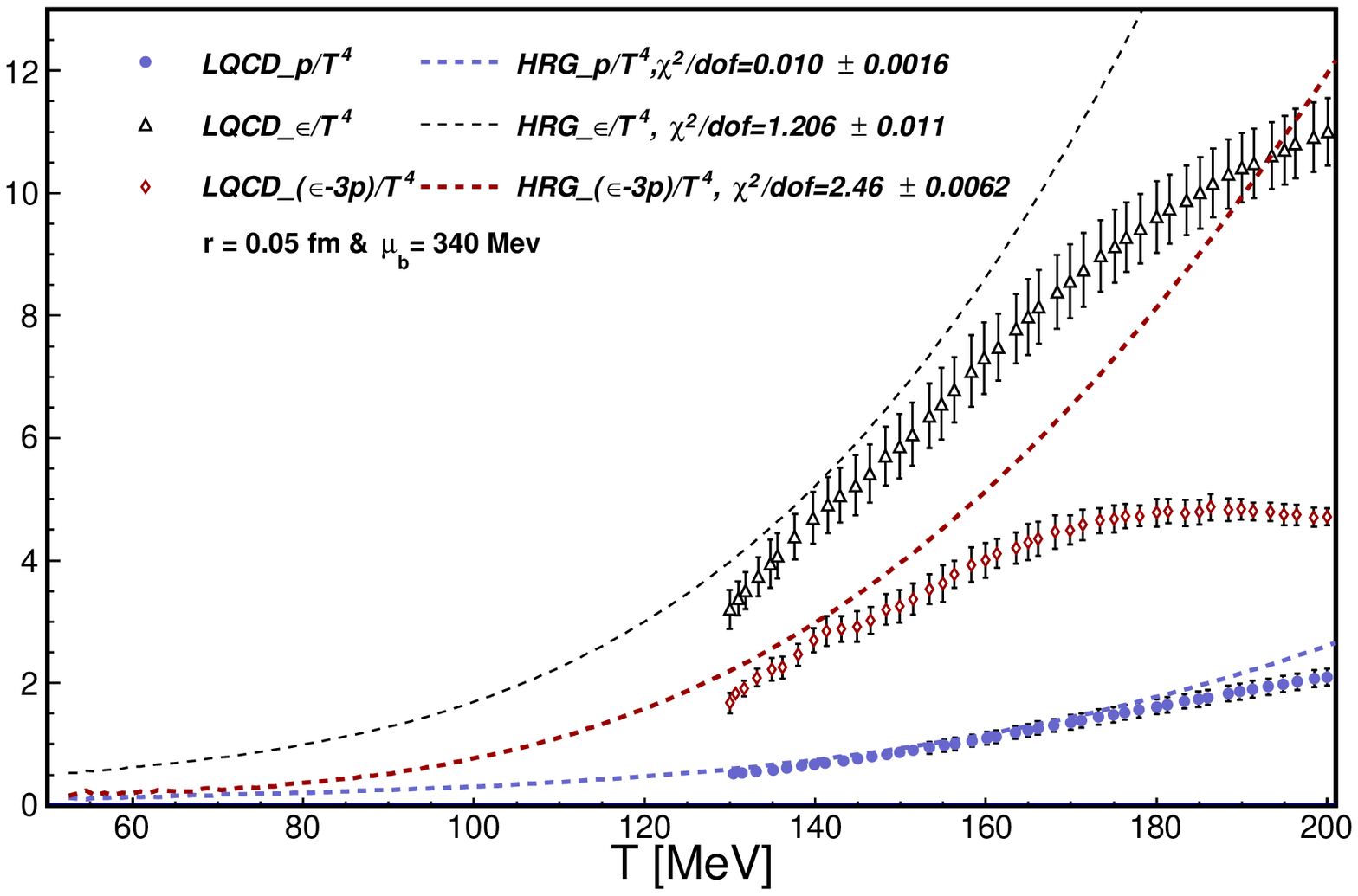} 
\includegraphics[width=8.25cm]{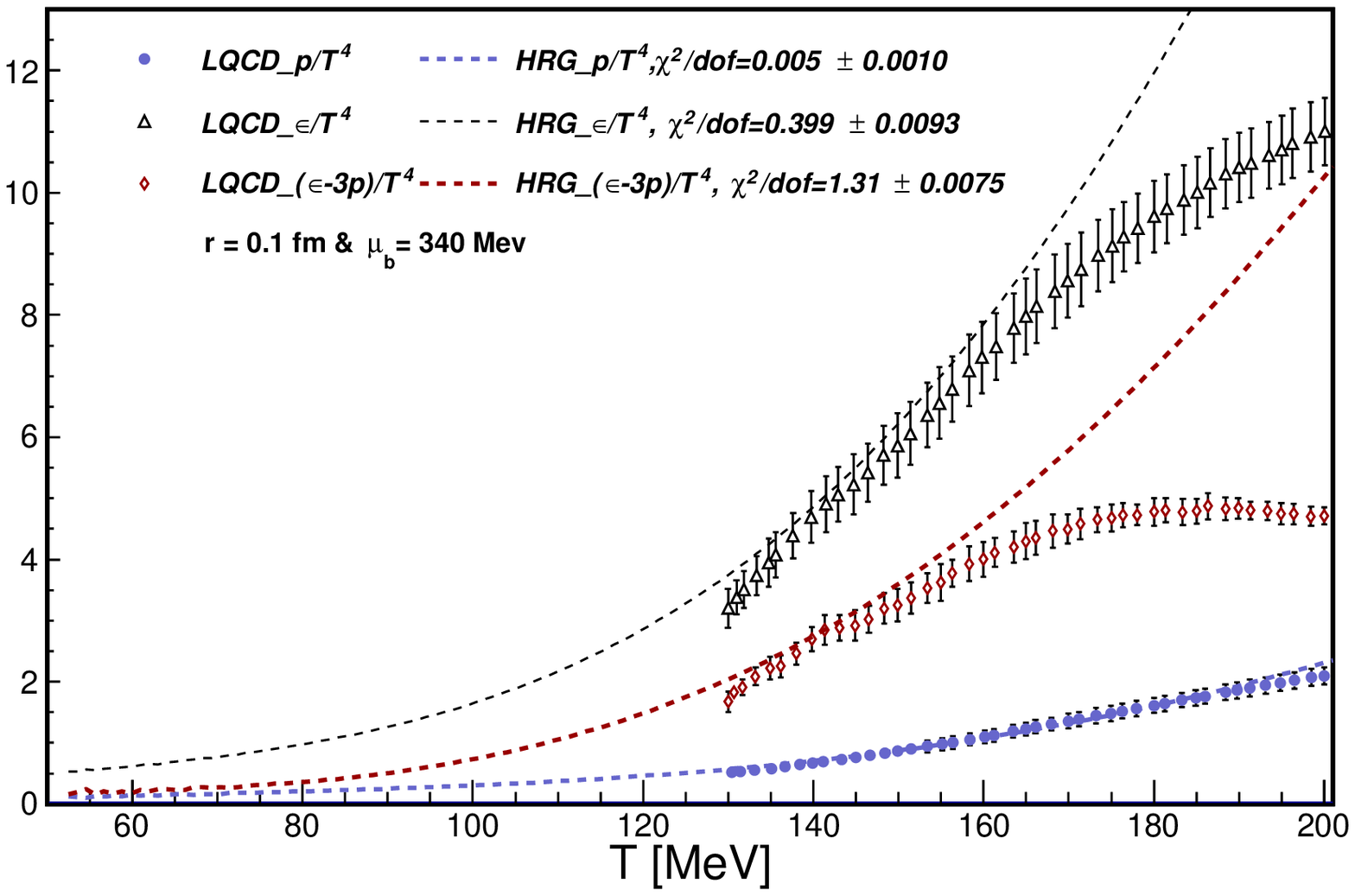} 
\includegraphics[width=8.25cm]{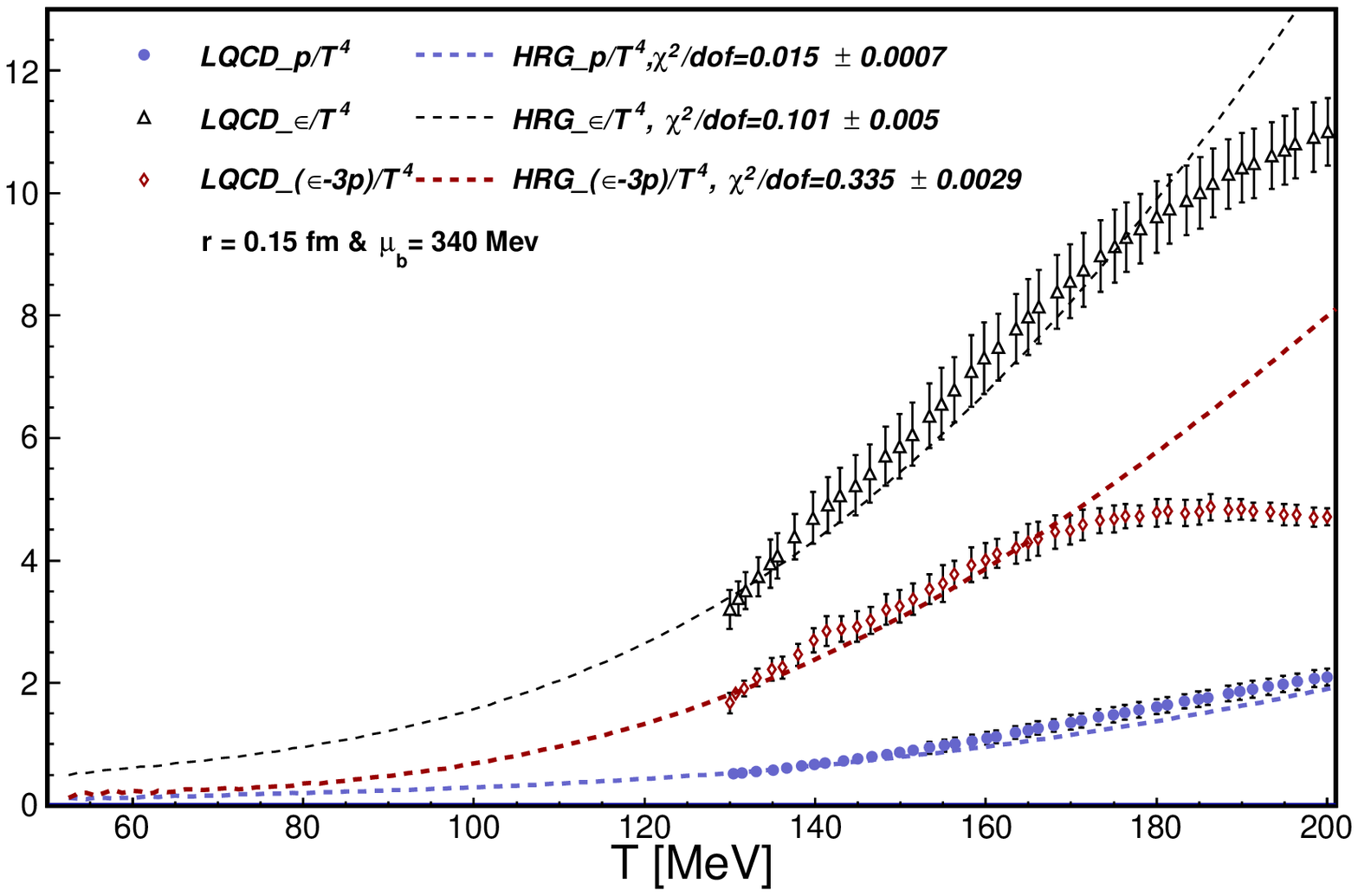}
\caption{The same as in Fig. \ref{fig:one} but at $\mu_{b}=340~$MeV.}
\label{fig:three}
\end{figure}

For the case of $\mu_{b}=340$ MeV, see Fig. \ref{fig:three}, the only interesting observation is that for $r=0.15~$fm, the model data well below and in the vicinity of $T_{c}$ and up to $T \backsimeq$ 170 MeV significantly approaches the corresponding lattice data. The data mismatch then diverges for higher temperatures.  

\begin{figure}[!htb]
\includegraphics[width=8.25cm]{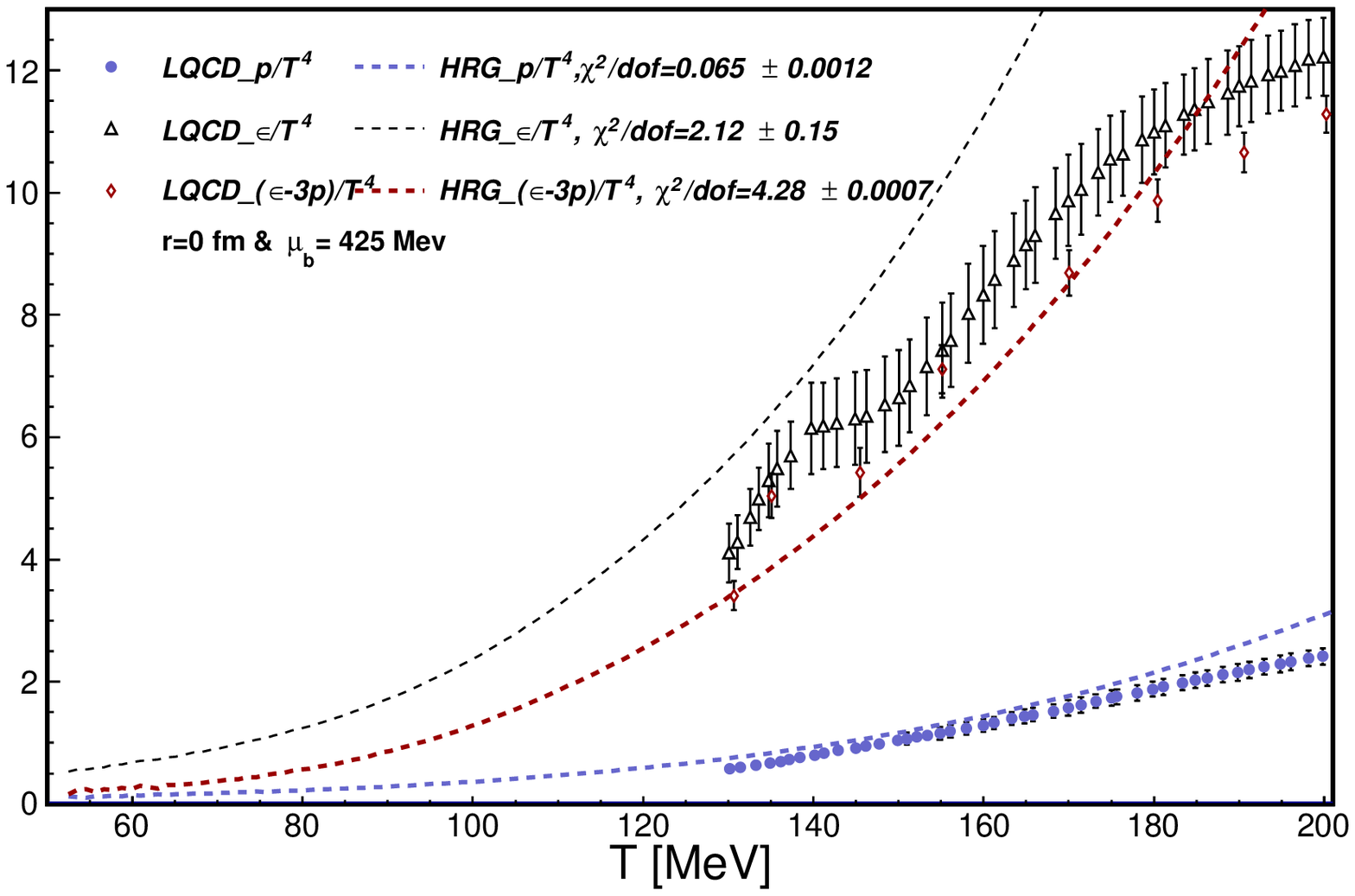}
\includegraphics[width=8.25cm]{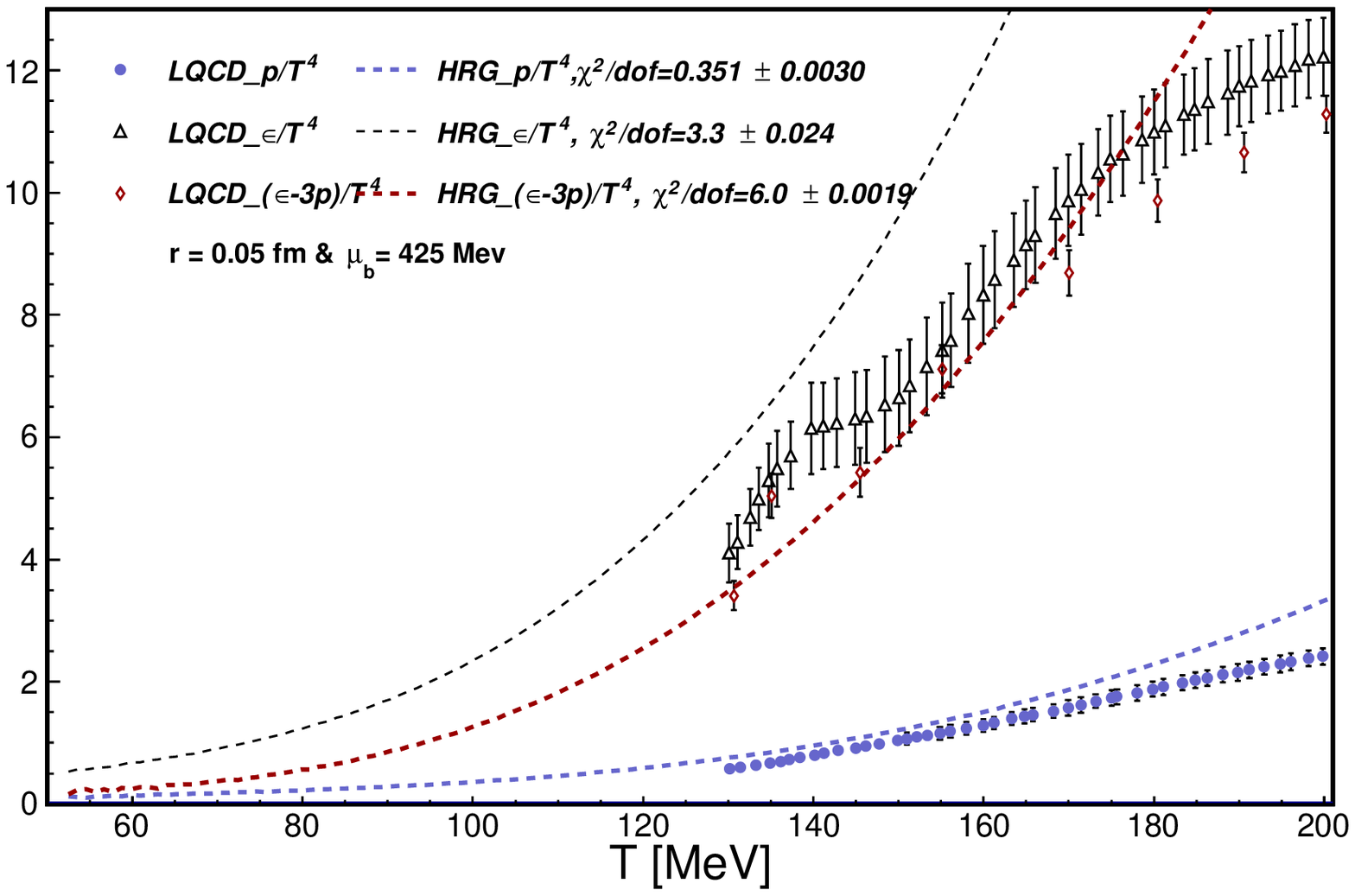} 
\includegraphics[width=8.25cm]{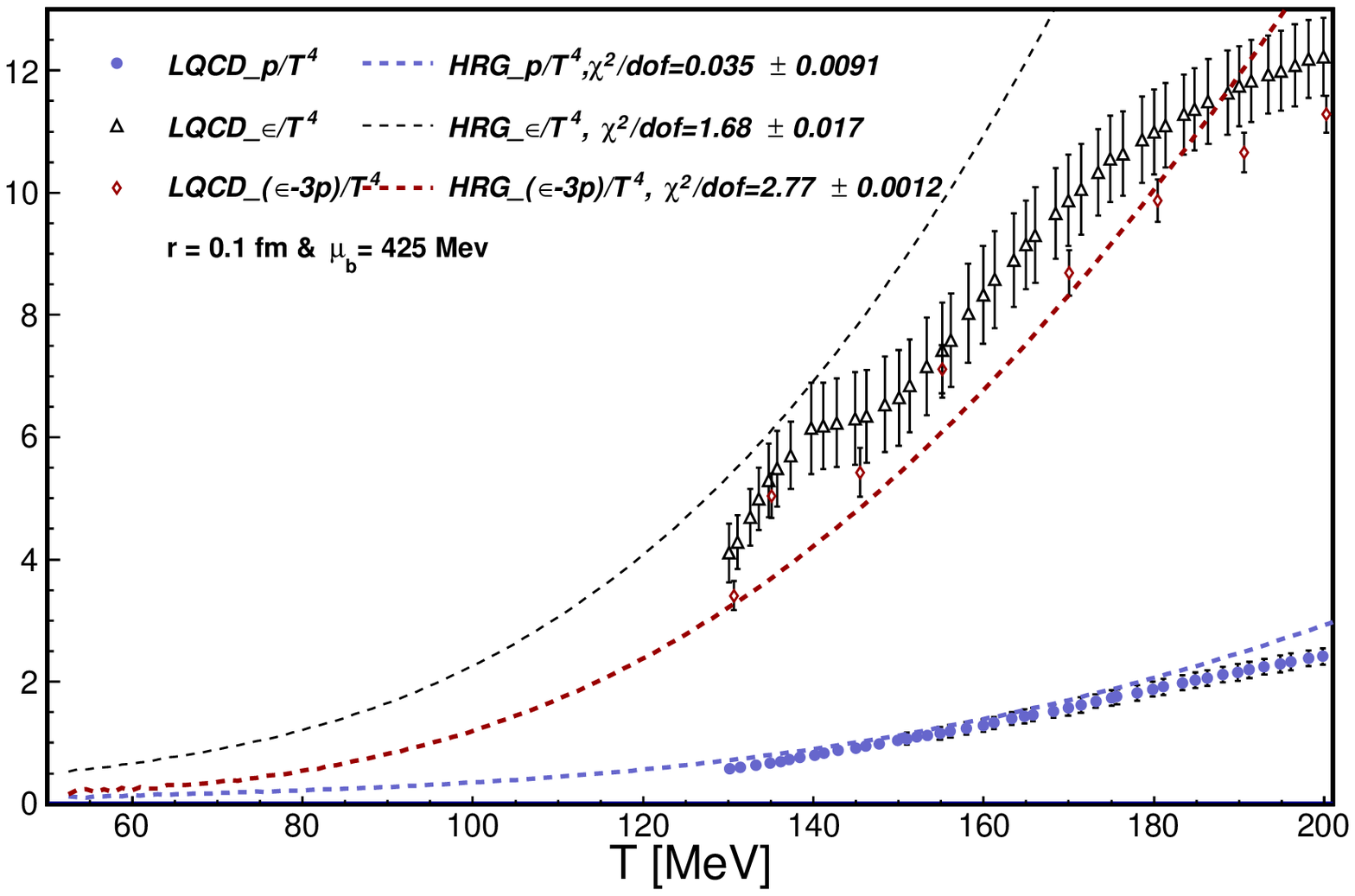} 
\includegraphics[width=8.25cm]{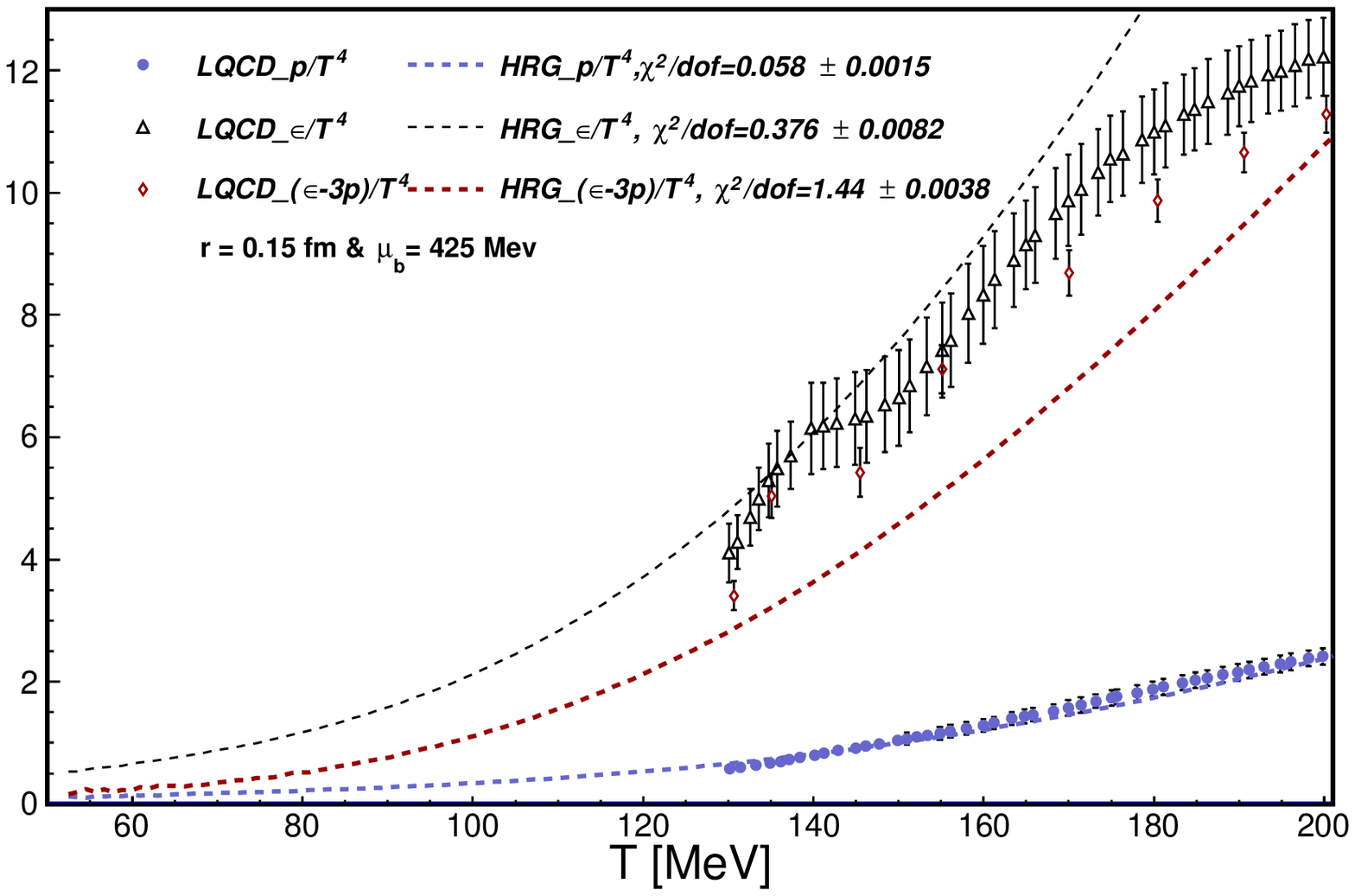}
\caption{The same as in Fig. \ref{fig:one} but at $\mu_{b}=425~$MeV.}
\label{fig:four}
\end{figure}

For the case of $\mu_{b}=425$ MeV, Fig. \ref{fig:four}, the data mismatch is generally too large to suggest any plausible correlation at any of values of $r$ and for all temperatures of interest. 

\begin{table}[htb]
\begin{tabular}{|c|c|c|c|c|}
\hline 
 $\mu_{b} $ MeV & $ r $ fm & $\chi^{2}$/dof for $P/T^{4}$ & $\chi^{2}$/dof for $\left(\rho-3P\right)/T^{4}$ & $\chi^{2}$/dof for  $\rho/T^{4}$ \\ 
\hline 
  & 0 & $0.020 \pm 0.00065$ & $0.061 \pm 0.0025$ & $0.080 \pm 0.0034$ \\ 
\cline{2-5}
  & 0.05 & $0.013 \pm 0.00037$ & $0.097 \pm 0.0011$ &  $0.089 \pm 0.0012$ \\ 
\cline{2-5}
0  & 0.1 & $0.009 \pm 0.00051 $& $0.091 \pm 0.0021$ & $0.174 \pm 0.0033$ \\ 
\cline{2-5}
  &0.15  &$0.102 \pm 0.0010$  & $0.024 \pm 0.0018$ & $0.370 \pm 0.0052$ \\ 
\hline 
  & 0 & $0.002 \pm 0.0002$ & $0.3698 \pm 0.0050$  & $0.036 \pm 0.0010$ \\ 
\cline{2-5}
  & 0.05 & $0.008 \pm 0.00041$ & $0.3190 \pm 0.0045$ & $0.062 \pm 0.002 $\\ 
\cline{2-5}
170  & 0.1 & $0.010 \pm 0.00062$ & $0.1452 \pm 0.0011$  & $0.075 \pm 0.0035$ \\ 
\cline{2-5}
  &0.15  & $0.028 \pm 0.0009 $& $0.069 \pm 0.0017 $& $0.145 \pm 0.0051 $\\ 
\hline 
  & 0 & $0.020 \pm 0.0014 $& $0.853 \pm 0.0028 $& $0.241 \pm 0.0040$ \\ 
\cline{2-5}
  & 0.05 & $0.010 \pm 0.0016$ & $2.46 \pm 0.0062$ & $1.206 \pm 0.011$ \\ 
\cline{2-5}
340  & 0.1 & $0.005 \pm 0.0010$ & $1.31 \pm 0.0075$ &  $0.399 \pm 0.0093 $\\ 
\cline{2-5}
  & 0.15  &$0.015 \pm 0.0007$ & $0.335 \pm 0.0029$  & $0.101 \pm 0.005$ \\ 
\hline 
  & 0 & $ 0.065 \pm 0.0012 $ & $ 2.12 \pm 0.15 $ & $ 4.28 \pm 0.0007 $  \\
\cline{2-5}
  & 0.05 & $0.351 \pm 0.0030$ & $6.0 \pm 0.0019$ &  $3.3 \pm 0.024$ \\ 
\cline{2-5}
425  & 0.1 & $0.035 \pm 0.0091$  & $2.77 \pm 0.0012 $&  $1.68 \pm 0.017$ \\ 
\cline{2-5}
  &0.15  & $0.058 \pm 0.0015$ & $1.44 \pm 0.0038$ & $0.376 \pm 0.0082$ \\ 
\hline 
\end{tabular}
\caption{$\chi^{2}$/dof statistic for the normalized pressure $P/T^{4}$, trace anomaly $\left(\rho-3P\right)/T^{4}$, and normalized energy density $\rho/T^{4}$ calculated in our statistically corrected hadron resonance gas (HRG) model confronted to the corresponding lattice data taken from ref. \cite{bazavov2017qcd,bazavov2014equation}, at four values of baryon chemical potential $\mu_{b}=0,170,340$, and $425~$MeV.}  
\label{tab1}
\end {table}

\section{Conclusions}
\label{sec:Cncls}

We confronted a novel statistical correction of the HRG model with recent lattice data  \cite{bazavov2017qcd,bazavov2014equation}. All our model calculations considered in this study doesn't seem to satisfactorily mimic the corresponding lattice data in the full temperature range under investigation, $T \in [130, 200~$MeV$]$. However, the best matching occurs locally in the vicinity of $T_{c} $ in the range $T \in [140, 170~$MeV$]$ for the case of $\mu_{b}=170~$MeV at zero and $0.05~fm$ correlation radii, $r$, respectively. Another remarkable matching between our model data with the corresponding lattice data occurs for the the case of $\mu_{b}=340~$MeV, at $0.1$ and $0.15~$fm correlation radii, respectively for temperatures $T \lesssim T_{c}$ and up to $T \backsimeq 170~$MeV. In the lower temperature range, $T \in [130, 160~$MeV$]$, most of the cases investigated in this research show reasonable match with the corresponding lattice data for different correlation lengths except for the case in which $\mu_{b}=425~$MeV where no good fitting is observed for any correlation length.

%
\bibliographystyle{aip}
\bibliography{statisticalcorrection2}

\end{document}